\documentclass[aps,preprint,amsmath,superscriptaddress]{revtex4}

\usepackage{dcolumn}
\usepackage{bm}
\usepackage{graphicx}
\usepackage{color}
\usepackage{subfigure}
\usepackage{latexsym}
\usepackage{amsthm}
\usepackage{amssymb}
\usepackage{multirow}
\usepackage[english]{babel}
\usepackage{hyperref}
\usepackage{cleveref}
\begin{document}
\def\be{\begin{equation}}
\def\ee{\end{equation}}

\def\bc{\begin{center}} 
\def\ec{\end{center}}
\def\bea{\begin{eqnarray}}
\def\eea{\end{eqnarray}}
\newcommand{\avg}[1]{\langle{#1}\rangle}

\newcommand{\Avg}[1]{\left\langle{#1}\right\rangle}

\title{Sparse power-law network model  for reliable  statistical  predictions based on sampled data}

\author{A. P. Kartun-Giles}
\affiliation{School of Mathematical Sciences, Queen Mary University of London, London, United Kingdom}
\author{D. Krioukov}
\affiliation{Departments of Physics, Mathematics, and Electrical \& Computer Engineering, Northeastern University, Boston,  United States}
\author{J. P. Gleeson}
\affiliation{MACSI, Department of Mathematics and Statistics, University of Limerick, Limerick, Ireland}
\author{Y. Moreno}
\affiliation{Institute for Biocomputation and Physics of Complex Systems (BIFI), University of Zaragoza, Zaragoza, Spain}
\affiliation{Department of Theoretical Physics, Faculty of Sciences, University of Zaragoza, Zaragoza, Spain}
\affiliation{ISI Foundation, Turin, Italy}
\affiliation{Complexity Science Hub Vienna, Vienna, Austria}
\author{G. Bianconi}
\affiliation{School of Mathematical Sciences, Queen Mary University of London, London, United Kingdom}
\begin{abstract}
A projective network model is a model that enables predictions to be made based on a subsample of the network data, with the predictions remaining unchanged if a larger sample is taken into consideration. An exchangeable model is a  model that does not depend on the order in which nodes are sampled. Despite a large variety of non-equilibrium (growing) and equilibrium (static) sparse complex network models that are widely used in network science,  how to reconcile sparseness (constant average degree) with the desired statistical properties of  projectivity and exchangeability is currently an outstanding scientific problem. Here we propose a  network process  with hidden variables which is  projective and can generate sparse power-law networks. Despite the model not being exchangeable, it can be  closely related to exchangeable uncorrelated networks as indicated by its information theory characterization and its network entropy. The use of the proposed network process as a null model is here tested on real data, indicating that the model offers a promising avenue for statistical network modelling.
\end{abstract}

\maketitle

\section{Introduction}

Network science \cite{Barabasi_book,Newman_book,Estrada_book,Vito_book} is one of the most rapidly advancing  scientific fields of investigation. The success of this field is deeply {rooted} in its interdisciplinarity. {In fact,} network science characterizes the underlying structure and dynamics  of complex system{s} ranging from on-line social networks to molecular networks and the brain. Additionally{, } the theoretical tools and techniques used by network science are coming from different disciplines including statistical mechanics, statistics,  machine learning and computer science.

In the last twenty years {significant} attention has been addressed to modelling framework of complex networks. Since  {most real} networks from the Internet to molecular networks, are sparse, i.e. they have an average degree that does not depend on the network size, statistical mechanics models focus on modelling sparse networks. These statistical mechanics models can be divided between non-equilibrium growing network models \cite{BA,BB,Doro_a,Krapivsky,Bianconi_Fortunato,Krapivsky_mod,Emergent,NGF,Hyperbolic} such as the famous Barab\'asi-Albert model \cite{BA} and equilibrium models such as maximum entropy network ensembles \cite{Bianconi2007,Bianconi2009,Anand2009,Anand2010,Sagarra,Frank} {including Exponential} Random Networks \cite{Snijders,Newman_Park,Anand2009,Anand2010, garlaschelli2008maximum} and block models \cite{Tiago1,Tiago2}. The non-equilibrium growing network models  have the power to {explain} the fundamental mechanisms giving rise to emergent properties such as scale-free distributions \cite{BA,Doro_a,BB,Krapivsky}, degree correlations \cite{BB}, communities \cite{Bianconi_Fortunato,Krapivsky_mod,Emergent} and network geometry \cite{Emergent,NGF,Hyperbolic}.
On the contrary{,} maximum network ensembles constitute the least biased models satisfying a given set of constraints. These models are not explanatory but constitute the ideal null hypothesis to which real network{s} can be compared.

Recently the need to formulate reliable statistical model{s} is {receiving} significant attention \cite{Airoldi}. A reliable statistical model will include projectivity and exchangeability ~\cite{kallenberg2002foundations,shalizi2013consistency,shalizi2017,aldous1981representations,diaconis2008graph}. The projectivity of the statistical network model guarantees  that the conclusions reached by considering a subsample of the data are consistent with the ones that can be drawn starting from a larger sample of the data. The exchangeability of the nodes implies that  the probability of a network does not depend on the specific labels of the nodes. However{,} how to reconcile these statistical requirements with the sparseness of the networks, i.e.{,} a average degree {that is independent of} the network size, constitute{s} a major impasse of network modelling. For instance it has been shown that random uncorrelated networks  are only projective if the average degree $\avg{k}$ of the network increases linearly  with the network size $N$, i.e.{,} if the network is maximally dense
and $\avg{k}=O(N)$ \cite{diaconis2008graph,shalizi2013consistency,krioukov2013duality}.

In physical terms the desired projective and exchangeable network process mimicking the subsequent sampling of an increasing portion of the network is a modelling framework that goes beyond the traditional statistical mechanics division between equilibrium and non-equilibrium network modelling approaches. This observation reinforces the {belief} that actually combining these two properties might be not an easy task.

Already several works have addressed this problem \cite{borgs2014lp,caron2017sparse,veitch2015class,borgs2016sparse,crane2016edge,cai2016edge,janson2017edge}, using different approaches such as relaxing the condition $\avg{k}=O(N)$ but  always characterizing models with average degree diverging with the network size $N$,  considering edge {exchangeable} models or alternatively using an embedding space as a basic mechanism to combine sparsity with projectivity and exchangeability \cite{krioukov2013duality,hoorn2017sparse}.

Here we propose a network process describing a network evolution mimicking the sampling of a network by subsequently expanding the nodes set. Each node is assigned an hidden variable from a hidden variable distribution. This distribution is the key quantity determining the properties of the network process. If the hidden variable is power-law {distributed} and the network is sufficiently sparse, the degree distribution displays a power-law tail with the same power-law exponent {as} the hidden variable distribution.

This model  is a projective network process but it is not exchangeable. Nevertheless  this non-equilibrium network model can be directly related to an equilibrium uncorrelated network ensemble in the sparse regime.
In fact{,} by permuting the order in which nodes are sampled it is possible to  calculate the probability that two nodes are connected given  their corresponding hidden variables.
This connection probability is equal to the connection  probability in an uncorrelated exchangeable network ensemble in which  the hidden variable of each  node is identified {with half} of  its expected degree.
The "proximity" between the network process and the uncorrelated network ensemble is here quantified by using information theory tools and comparing the entropy of the two models. In {particular,}  we use the entropy of the two network models \cite{Bianconi2007,Bianconi2009,Anand2009,Anand2010,PNAS} to evaluate the difference in the information content of the two models{,} finding that the two models have small relative entropy difference.

Finally we {study} how well the proposed model can be used as a null model for real power-law network datasets. To this end we {identify} the hidden variable of each node with half  of its observed degree and we {run} the model by adding the nodes in the network according to a random permutation of the {nodes'} labels.
The degree distribution of the real dataset and the degree distribution of the simulation results are in good agreement when starting from power-law networks, and the agreement remains good if the network is grown by only considering a subsample of the nodes of the real data.
We {also compare} the correlations {of} the real dataset with the correlations of the simulation results {to show} that the simulations are able to generate only weak correlations of the degrees. Therefore a more refined model should be formulated to capture this additional network property.

The paper is structured as described in the following. In Sec. 2 we introduce the definition of the {desired statistical} properties of network models: {\em projectivity} and {\it exchangeability}. In Sec. 3 we discuss major examples of sparse {network} models (the Barab\'asi-Albert model and the uncorrelated network ensembles) and characterize them with respect to the {properties} of projectivity and exchangeability. In Sec. 4 we present an account of the difficulties in combining projectivity and exchangeability with the sparseness of networks and we give a brief review of the approaches investigated in the recent literature on the subject. In Sec. 5 we present a network process mimicking a network sampling process. We characterize its structural and dynamical properties relating this non-equilibrum model to equilibrium uncorrelated network ensembles{,} and we characterize its statistical properties. In Sec. 6 we show the possible use  of the proposed network process  as a null model for modelling real power-law network datasets. Finally in Sec. 7 we give the conclusions.

\section{Statistical terms}

Projectivity and exchangeability are two very basic and very natural statistical requirements for {reliable} statistical  network models.  In physical terms, projectivity is directly related to the principle of locality, while exchangeability is related to symmetry. In this section, we first discuss projectivity and exchangebility to make clear that they really are ``must-have''s in any statistically { useful} network model, while in the next two sections we will comment on difficulties in combining them both in models of sparse networks, i.e., having average degree independent of the network size $N$ \cite{sparse_note}.
{  While projectivity and exchangeability  are desired properties of  statistically reliable network models,  the relevance and of these requirements for any realistic network model is a subject of scientific debate  (see for instance contribution of Karthik Bharath in the discussion of the F. Caron and E. Fox paper \cite{caron2017sparse}).In fact it is often observed  that most real networks can hardly be exchangeable. Indeed, in a vast majority of real networks nodes are labelled with labels related to some rich metadata and a random permutation of the nodes labels would result in a different network whose probability to be produced by the same stochastic process that produces the real network is certainly not expected to be equal to the probability with which it generates the real network. }

In order to investigate the properties of reliable statistical models we consider a  network process mimicking the subsequent sampling a network by expanding the  set of sampled nodes  and detecting all the interactions among this set of nodes.

To this end we consider  a set of networks $\{G_t\}_{t=1,2,\ldots}$ with $G_t=(V_t,E_t)$ and increasing network size $N_t=|V_t|=t$. The sequence of networks defines a network process, i.e. $G_t=(V_{t},E_t)$ is an induced subgraph of the network $G_{t'}=(V_{t'},E_{t'})$ for all $t<t'$  with node set  $V_{t}\subset V_{t'}$  if $t<t'$. We label the nodes in order of their appearance in the network such that
\bea
V_t=\{1,2,\ldots, t\}.
\eea
and assign a probability $P(G_t)$ to each  network $G_t$.

\subsection{Projectivity}

Given the set of networks  $\{G_t\}_{t=1,2,\ldots}$ {\it projectivity} implies that  the statistical properties of the network $G_{t}$  are directly related to the statistical properties of the network $G_{t'}$ with $t'>t$ by a proper marginalization of the probability of  the network $G_{t'}$ over its subgraph  $G_{t}$.

By definition~\cite{kallenberg2002foundations,shalizi2013consistency}, a projective network model is a model that attributes a given probability $P(G_t)$ to each network $G_t$ of the sequence,  such that $$P(\pi_{t',t}(G_{t'}))=P(G_t),$$ where the projective map $\pi_{t',t}$  maps networks $G_{t'}$ of a larger size $t'>t$ to their subgraph $G_t$ of a smaller size $t$.

In other words this means that one can first generate a larger graph $G_{t'}$ using the model, then reduce its size to $t$ by throwing out some $t'-t$ nodes according to the projective map specification, and the probability with which the resulting graph $G_t$ is generated using this two-step procedure will be the same as if graph $G_t$ was generated by the model directly.

\subsection{Exchangeability}

Exchangeability implies that the order in which two nodes are observed or labelled is not important.
 Specifically, a network model is exchangeable if, by definition~\cite{aldous1981representations,diaconis2008graph},
 the probability $P(G_t)$ of a network $G_t=(V_t,E_t)$ is independent on the nodes labels, i.e.
 \bea
 P(G_t)=P(\tilde{G}_t)
 \eea 
 where $\tilde{G}_t$
 is any  network isomorphic to the network $G_t$, i.e. it is any network 
 obtained from the network $G_t$ by permuting  the nodes  labels $\{i\}_{i=1,2,\ldots,N}$ 
 according to the permutation $\bm{\sigma}$.
If a network model is exchangeable it follows that the marginal 
 the probability $p_{ij}$ of  the generic  link between node $i$ and node $j$ is unchanged if the node labels are permuted,
 implying that they are sampled in a different order, i.e.{,}
\bea
p_{ij}=p_{\sigma(i),\sigma(j)}{.}
\eea
Therefore exchangeability  enforces  the {\it symmetry} of the model with respect to the group of graph isomorphisms.

\section{Characterization of relevant sparse network models from the statistical perspective}

In this section we investigate major examples of non-equilibrium (growing) network models and
equilibrium (static) network models widely used to model sparse complex networks.
In particular we discuss the Barab\'asi-Albert model \cite{BA} and the uncorrelated network ensembles from the statistical perspective.
This discussion will reveal that neither of these two very popular frameworks {for modelling} sparse complex networks display both projectivity and exchangeability{,} indicating the difficulties in combining these properties with the sparseness of the networks.

\subsection{Barab\'asi Albert model}
{The} Barab\'asi-Albert model {begins with} an initial finite network {and} at each time $t$ a new node enters in the network and {is} connected to the network by establishing $m$ new links.
Each of these links connect the new node to a node $i$ with degree $k_i$ chosen with probability
\bea
\tilde{\Pi}_i=\frac{k_i}{\sum_{i'}k_{i'}}.
\eea
This probability  enforces {\it preferential attachment}, i.e.{, allows} nodes with higher degree to {more rapidly acquire} new links.

The Barab\'asi-Albert model describes a model that is projective, because as the network grows the network $G_t$ obtained at time $t$ is an induced subgraph of the network $G_{t'}$ obtained at a later time $t'>t$.
However the Barab\'asi-Albert model is not exchangeable.
The fact that the network is not exchangeable is  revealed for instance by the expression {for} the average number of links $k_i(t,t_i)$ of a node $i$ arrived in the network at time $t_i$,
\bea
k_i(t,t_i)=m\left(\frac{t}{t_i}\right)^{1/2}.
\eea
This expression explicitly indicates  that  the older nodes are {statistically} different from the younger nodes, and their degree is much larger than {that} of younger nodes.
Additionally it is possible to observe that the model is not exchangeable because the order of the addition of the  nodes, i.e.{,} their time of arrival in the network, is the key property that determines the connection probability \cite{Bianconi_Ising}, i.e.{,}
\bea
p_{ij}\simeq\frac{m}{2}\frac{1}{\sqrt{t_i t_j}}.
\eea
Nevertheless  we observe the interesting property that  for this model the connection probability $p_{ij}$ between node $i$ and node $j$ can be also expressed as
\bea
p_{ij}\simeq \frac{k_i(t,t_i)k_j(t,t_j)}{\sum_{i'}k_{i'}(t,t_{i'})},
\eea
indicating that actually, although the network process has different statistical properties than the uncorrelated network with the same degree distribution, the expected degree correlations are weak. The relation between the BA model and the uncorrelated network ensemble with the same degree distribution is investigated in detail using information {theoretic} tools in Ref. \cite{entropy_rate}.

\subsection{Uncorrelated Network Ensembles}

The Barab\'asi-Albert model is projective but not exchangeable. On the contrary the widely used uncorrelated network ensembles are exchangeable models but they are not  projective  in the sparse regime. In order to show this let us consider  an uncorrelated network model in which each node $i$  has an expected degree $\theta_i$, where the expected degree{s} of the nodes are consistent with a structural cutoff, i.e.
\bea
\theta_i\leq \sqrt{\avg{\theta}N}.
\eea
In this case the probability $p_{ij}$ of a link between node $i$ and node $j$ is given by
\bea
p_{ij}=\frac{\theta_i \theta_j}{\Avg{\theta}N},
\eea
and therefore it only depends on the expected degree{s} $\theta_i$ and $\theta_j$ of the nodes $i$ and $j$ and not on the order in which node $i$ and node $j$ have been sampled.
The model is therefore exchangeable as long as we consider the {simultaneous} permutation of the node labels and the expected degrees of the nodes.
However if we consider a large sample of the network with $N'>N$ nodes, we see that the model is projective if and only if it is also dense, with the number of links scaling as $L=O( N^2)$.
In fact if we assume that in the larger sample the expected degrees of nodes $i$ and $j$ are given by $\theta_i^{\prime}$ {and $\theta_j^{\prime}$}, the probability that node $i$ and node $j$ are {connected} in the larger network models including $N^{\prime}$ nodes is
\bea
p_{ij}^{\prime}=\frac{\theta_i^{\prime} \theta_j^{\prime}}{\Avg{\theta^{\prime}}N^{\prime} }
\eea
If we impose projectivity, i.e.
\bea
p_{ij}=p_{ij}^{\prime}
\eea
for $i,j \leq N$, and we assume that   the number of nodes $N^{\prime}>N$ can be written as
\bea
N^{\prime}= zN,
\eea
it is easy to see   that we should also have
\bea
\theta_i^{\prime} &=& z\theta_i, \nonumber \\
\avg{\theta_i^{\prime}}&=&z\avg{\theta}.
\eea
Therefore to guarantee projectivity  the expected degree of each node should grow linearly with the network size, resulting in a dense network with  the total number of links $L$ scaling with the network size $N$ as $L={O}(N^2)$.
This implies that  the random network $\mathbb{G}(N,p)$ with $p$ independent {of} $N$ is an exchangeable model whereas the Poisson random network $\mathbb{G}(N,p)$ with $p=\frac{z}{N}$ and $z$ independent {of} $N$ is not exchangeable.
In fact one cannot throw out $N'-N$ nodes from a network of size $N'$ produced by $\mathbb{G}(N',z/N')$, and hope that the resulting network  will have the same probability as in $\mathbb{G}(N,z/{N})$, simply because the links in the  $\mathbb{G}(N',z/N')$ and $\mathbb{G}(N,z/N)$ {ensembles} exist with different probabilities $z/N'$ and $z/N$ that depend on the graph size $N$. Alternatively, if one attempts to formulate $\mathbb{G}(N,z/N)$ as a growing model, then since the edge existence probability depends on $N$, {the} addition of a new node affects the probability of existence of edges in the existing network. Since this probability is a decreasing function of $N$ ($z/N$), upon {the} addition of a new node all the existing edges must be removed with some probability ($1/N$). In other words, in such a growing model new node additions must necessarily affect the existing network structure.

\section{Impasse with sparsity}

Surprisingly, combining  projectivity and exchangeability {with the additional constraint of} sparsity, i.e. the requirement that the average degree of the sampled networks is independent {of} the network size, has been a major impasse. If we exclude spatially embedded networks \cite{krioukov2013duality}, to the best of our knowledge there exists no model of sparse networks that would be both projective and exchangeable at the same time. This situation is in stark contrast with {the case} of dense graphs. Dense graphs are known to have well-defined thermodynamic limits known {as} graphons, and any graphon-based network model is both exchangeable and projective~\cite{diaconis2008graph}.

The thermodynamic limits of sparse graphs are at present quite poorly understood, which appears to be one of the reasons behind the mentioned impasse. Several attempts have been made to understand the limits of sparse graphs, including, for example, sparse $L^p$ graphons~\cite{borgs2014lp}, which are not projective, or stretched graphons a.k.a.\ graphexes~\cite{caron2017sparse,veitch2015class,borgs2016sparse}. In the latter case, graphs are sparse, exchangeable and projective, but with two major caveats:
\begin{itemize}\item[1)]~the average degree cannot be constant, it must diverge with $N$ (but possibly slower than linearly),
\item[2)]~exchangeability is completely redefined: it is not with respect to node labels $1,\ldots,N$, but with respect to artificial labels which are positive real numbers.
\end{itemize}
Another class of attempts suggests to completely give up on the node label exchangeability requirement, and to consider edge exchangeability instead, e.g., using variations of Pitman–Yor processes~\cite{crane2016edge,cai2016edge,janson2017edge}. It remains unclear at present whether these developments imply that too many network models that were found to be quite useful in practice and that do use node labels $1,\ldots,N$, are statistically hopeless. It seems more likely that further research is needed to understand and resolve this projectivity vs.\ exchangeability impasse in sparse network models.

\subsection{Proposed solution of the impasse based on network geometry}

In~\cite{krioukov2013duality} it was shown that a generic network model is projective if the probability of edge existence, i.e. the connection probability, does not depend on the network size $N$. In fact if the connection probability does depend on $N$, then, {the} addition of new {nodes} to the existing network in the growing formulation of the {model necessarily} affects the existing network structure  and the network cannot be projective.

In order to formulate network models in which the connection probability does not depend on the network size $N$, embedding networks in space can turn out to be very useful.
In fact   spatially embedded networks can combine  projectivity with a constant average degree~\cite{krioukov2013duality}  as their spatial embedding  ensures projectivity when the connection probability is {\it local} and nodes connect typically to nodes that are spatially close.  For instance if the nodes are uniformly distributed in $\mathbb{R}^2$  and each node connects only to the nodes with a constant radius $r_0$, by sampling the network by progressively expanding the spatial region of interest we can build a projective model with constant average degree.
This is clearly a realistic scenario in most real networks as it unlikely that a local event in a spatial network causes a global change in the network. For instance in the Internet, the appearance of a new customer of a local Internet provider in Bolivia {cannot} lead to immediate severance of customers by a local Internet provider in Bhutan.

It turns out that {models} that are not explicitly constructed {from spatial embeddings can also} be analysed using geometrical arguments{, hence shedding} light on their statistical properties.
In this vein, it was recently shown that the hypersoft configuration model, {which} defines maximum-entropy random graphs with a given degree distribution, is sparse and {\it either} exchangeable {\it or} projective~\cite{hoorn2017sparse}. Both sparsity and exchangeability definitions are traditional in the model, i.e., the average degree is constant and exchangeability is with respect to labels $1,\ldots,N$, so that the only caveats are in ``either-or'' and also in that this ``either-or'' is achieved only for specific degree distributions (power law with exponent $\gamma=3$ in~\cite{hoorn2017sparse}).

 In the {\it exchangeable} equilibrium formulation of the model, nodes are points sprinkled at random onto an interval $A_N$ of an $N$-dependent length $L_N$, where $L_N$ is a growing function of $N$, according to a non-uniform point density (if this point density is exponential, then the resulting degree distribution is a power law), and then all pairs of points/nodes $i$ and $j$, $j>i=1,\ldots,N$, at sprinkled coordinates $x_i$ and $x_j$ are connected by an edge with the entropy-maximizing Fermi-Dirac connection probability
 \bea
 p(x_i,x_j)=\frac{1}{e^{x_i+x_j}+1}
 \eea
  that does not depend on the network size $N$.

In the {\it projective} growing formulation of the same model, the interval $A_N$ grows with $N$, its length growing according to $L_n$, new node $N+1$ appears in the interval increment $A_{N+1} \setminus A_N$ of length $L_{N+1}-L_N$, and then connects to existing nodes with the same connection probability as in the exchangeable formulation.

The difficulty of combining projectivity and exchangeability is evident in this example: in the exchangeable formulation, node labels $i$ are random and uncorrelated with their coordinates $x_i$, while in the projective formulation, nodes are labeled in the increasing order of their coordinates: $i<j \leftrightarrow x_i<x_j$. If nodes are labeled this way, then the projective map $\pi_{N',N}$ trivially throws out nodes with labels $N+1,\ldots,N'$, and the resulting graph satisfies the projectivity requirement since the connection probability does not depend on $N$, and since the remaining $N$ nodes lie in $A_N$. If the node labels are random however, as they are in the exchangeable formulation, then it remains unclear if even an asymptotically correct projective map can be constructed.

\section{Statistical mechanics model with hidden variables}

Our goal is here to reconcile sparseness with a reliable statistical modelling framework without assuming the existence of an embedding geometrical space.
In this endeavour we will define  a projective network process yielding a sequence of networks growing by the subsequent addition of nodes and links.
To each node $i$ we associate {a hidden} variable  $\theta_i$ that is a proxy for the degree that the node will acquire in the model.
The statistical properties of the network model when we average over all the possible sequences determining the subsequent addition of the links obey scaling laws and reduce to the uncorrelated network model of any size $N$ in the sparse regime.

Although this model does not ultimately  reconcile sparseness with both exchageability and projectivity, we will see in Sec. $\ref{Stat}$ that it provides a very reliable null model for power-law networks also if only a subsample of the original network is considered.

\subsection{The model}
The model can be interpreted as a weighted growing network model where we allow multiedges.
In the model every node $i$  is assigned {a hidden} variable $\theta_i$ from a hidden variable distribution $\rho(\theta)$.\\
Starting at $t=1$ from a single isolated node,
at each time $t>1$ a new node $i$ is added to the network and draws $\kappa_i$ links to the existing nodes of the network, where $\kappa_i$ is chosen according to the Poisson distribution with average $\theta_i$, i.e.
\bea
\hat{P}(\kappa_i|\theta_i)=\frac{1}{\kappa_i!}\theta_i^{\kappa_i} e^{-\theta_i}.
\eea
Each new link is attached to a node $j$ already present in the network with probability
\bea
\Pi_j=\frac{\theta_j}{\sum_{r=1}^{t-1} \theta_r}.
\eea
Note that not all the new links will yield new connections because the nodes $i$ and $j$ might be already connected.
Additionally note that this model does not implement preferential attachment as the linking probability is only dependent on the externally attributed hidden variable $\theta_i$ and not to the dynamically acquired degree $k_i$. {Whenever} a new link connects node $i$ to an already connected node $j$ the multiedge between node $i$ and node $j$ is reinforced, i.e. the weight of the links between node $i$ and node $j$ increases by one.

Here and in the following we will indicate {by} ${\bf a}$ the adjacency matrix of the network, with $t_i$ the time at which node $i$ has been added to the network, with $k_i$ the node degree and with $s_i$ the node strength, i.e.{,} the sum of the weights of the links incident to node $i$.

\subsection{The strength of a node and its dependence on the hidden variable }
The hidden variable $\theta_i$ modulates the temporal evolution of the strength of the node $i$.
In fact in the mean-field approach \cite{BA,Doro_book,Barabasi_book}, since at each time {an} average of $\avg{\theta}$ links are added and reinforced, the average strength $s_i(t|t_i, \theta_i,\kappa_i)$ of node $i$ given the time $t_i$ of its arrival in the network, its hidden variable $\theta_i$ and its initial strength $\kappa_i$ obeys the equation
\bea
\frac{ds_i}{dt}=\avg{\theta}\frac{\theta_i}{\avg{\theta}t}=\frac{\theta_i}{t}
\eea
with initial condition $s_i(t_i|t_i, \theta_i,\kappa_i)=\kappa_i$.
{The solution of this equation is}
\bea
s_i(t|\theta_i,\kappa_i)=\theta_i\ln\left(\frac{t}{t_i}\right)+\kappa_i.
\eea
Therefore in this model the strength depends both on the time of arrival of the node in the  network and on its hidden variable.
If we average the strength over the nodes with the same hidden variable{ however,} we see that the average strength $\tilde{s}_i(\theta_i)$ of nodes with hidden variable $\theta_i$ is given  in the large network limit $t\gg 1$ by
\bea
\tilde{s}_i(\theta_i)=2\theta_i.
\eea
In fact we have
\bea
\Avg{\kappa_i|\theta_i}&=&\theta_i\nonumber \\
\tilde{s}_i(\theta_i)&=&\frac{1}{t}\int_{1}^t\theta_i \ln\left(\frac{t}{t_i}\right) dt_i+\Avg{\kappa_i|\theta_i}\nonumber \\
&=& 2 \theta_i+O\left(\frac{\ln t}{t}\right).
\eea
This implies that if we attribute to a node {a hidden} variable $\theta_i$ and we consider a set of models in which the time of arrival of node $i$ is taken {randomly, the strength of node $i$ is (on average over the different network models) determined  only by its hidden variable.}

\subsection{Strength distribution}

The strength distribution of the model is a convolution of exponentials.
To find the strength distribution we use the master equation approach \cite{Doro_book} under the assumption that the hidden variable distribution has a well defined average value $\avg{\theta}$. To this end we write the equation for $N^{t}_{\theta} (s)${,} the average number of nodes with hidden variable $\theta$ that have strength $s\geq 0$ at time $t${, as}
\bea
\frac{N^{t}_{\theta} (s)}{dt}=\avg{\theta}\Pi(\theta)N^{t}_{\theta} (s-1)[1-\delta(s,0)]-\avg{\theta}\Pi(\theta)N^{t}_{\theta} (s)+\rho(\theta)\hat{P}(\kappa=s|\theta),
\label{me}
\eea
where $\delta(x,y)$ indicates the Kronecker delta and where we {denote by} $\Pi(\theta)$ the probability that a node with hidden variable $\theta$ is attached to the new node arrived in the network at time $t$ by one of its connections, i.e.
\bea
\Pi(\theta)=\frac{\theta}{\sum_{\theta'} \theta' \sum_{s}N_{\theta'}^{t}(s)}\simeq \frac{\theta}{\avg{\theta}t}.
\eea
Given the continuous growth of the network asymptotically in time{, for} $t\gg 1$ it is possible to assume that
\bea
N^{t}_{\theta} (s)\simeq t P_{\theta}(s),
\eea
where $P_{\theta}(s)$ is the probability that a random node has strength $s$ and hidden variable $\theta$.
By inserting this asymptotic expression in the master equation  $(\ref{me})$ and solving for $P_{\theta}(s)$ we get
\bea
P_{\theta}(s)=\rho(\theta)\sum_{\kappa=0}^s\hat{P}(\kappa|\theta)\frac{1}{1+\theta}\left(\frac{\theta}{1+\theta}\right)^{s-\kappa}.
\eea
Therefore given the value of the hidden variable $\theta$ and the initial number of links $\kappa$ the strength distribution is exponential.
The overall strength distribution $P(s)$ of the model determining the probability that a random node has strength $s$ is given by the integral of $P_{\theta}(s)$ over all possible value of the hidden variable $\theta$, i.e.
\bea
P(s)=\int d\theta \rho(\theta)\sum_{\kappa=0}^s\hat{P}(\kappa|\theta)\frac{1}{1+\theta}\left(\frac{\theta}{1+\theta}\right)^{s-\kappa}.
\label{Ps}
\eea
This result reveals that the strength distribution can be different from the distribution of hidden variables. For instance if all the hidden variables are the same, the strength distribution will still allow for fluctuations of the strengths.
However for power-law hidden variable distributions
\bea
\rho(\theta)\simeq C\theta^{-\gamma}
\label{rhog}
\eea
the strength distribution has a power-law tail with the same exponent $\gamma$
\bea
P(s)\simeq \hat{C} s^{-\gamma}
\eea
for $s\gg 1$.
In fact, by inserting the explicit expression of $\hat{P}(\kappa|\theta)$ and of $\rho(\theta)$ in Eq. $(\ref{Ps})$ we get
\bea
P(s)=C\int d \theta \frac{\theta^{-\gamma}}{1+\theta}\sum_{\kappa=0}^s\frac{1}{\kappa!}\frac{\theta^s}{(\theta+1)^{s-k}} e^{-\theta}{.}
\eea
For $s\gg 1 $ we can approximate the sum over $\kappa$ with the infinite sum getting
\bea
P(s)=C\int d \theta \frac{\theta^{-\gamma}}{1+\theta}\left(\frac{\theta}{\theta+1}\right)^s e^{-1}\simeq \hat{C}s^{-\gamma}
\eea
where the last expression is valid {if} $s\gg 1$.
Therefore, although in general it is not true that the hidden variable distribution is the same as the strength distribution, in the case of power-law distributed hidden variables the strength distribution displays a power-law tail with the same exponent.  Note that this is valid for power-law exponents in the range $\gamma \in (2,3]$ but also in  the range $\gamma \in(1,2]$. Therefore in this case the hidden variables can be used to directly tune the strength distribution.

\subsection{Connection probability}

In this {section we derive} the expression for the connection probability {between} any two nodes.
Let us  consider the probability $P(a_{ij}=1|\theta_i,\theta_j, \kappa_j,t_j>t_i)$ that node $i$ is connected to node $j$, i.e.  $a_{ij}=1$ given the hidden variables of node $i$ and node $j$, their time of arrival with $t_j>t_i$ and the initial strength $\kappa_j$ of node $j$. This probability is one minus the probability that {all} of the initial links of node $j$ {do not} connect to node $i$, i.e.
\bea
P(a_{ij}=1|\theta_i,\theta_j\kappa_j,t_i<t_j)=1-\left(1-\frac{\theta_i}{\sum_r\theta_r}\right)^{\kappa_j}.
\eea
If we now average over the probability $\hat{P}(\kappa_j|\theta_j)$ we get the closed form expression
\bea
P(a_{ij}=1|\theta_i,\theta_j,t_j,t_i<t_j)=\sum_{\kappa_j}P(\kappa_j)\left[1-\left(1-\frac{\theta_i}{\sum_{r=1}^j\theta_r}\right)^{\kappa_j}\right]=\left[1-\exp\left(-\frac{\theta_i\theta_j}{\avg{\theta}t_j}\right)\right],
\label{conn_p}
\eea

where we have assumed that the average of the hidden variables $\Avg{\theta}$ is well defined.
Therefore we have found that the connection probability {between} two nodes depends both on the hidden variables and on their time of arrival in the network.
It follows that the model is not expected to be exchangeable, as this would require a connection probability independent {of} the time of arrival of the two nodes. However the fact that this connection probability does not {\it only} depend on the time of arrival of the nodes in the network (or the order in which they are sampled)
can be a useful {characteristic} of  a reliable statistical  model.

\subsection{Degree distribution in the sparse regime}

{Here} we derive the degree distribution of the model in the sparse regime, when we can assume that $p_{ij}\ll 1$. We will show that in this regime, each node has a Poisson degree distribution with an expected average degree $\overline{k}_i$ depending both on the value of its hidden variable and on the {time} of its arrival in the network.

The probability $P(k_i|\theta_i, t_i)$ that a node $i$ arrived in the network at time $t_i$ and{,} having  hidden variable $\theta_i${,} has degree $k_i$ can be calculated starting from the connection  probabilities  $p_{ij}$  given by Eq. $(\ref{conn_p})$.
Let us indicate with ${\bf a}_i=\{a_{ij}|j\in \{1,2,\ldots, N\}\}$ the elements of the adjacency matrix in the $i$-th row indicating the connections of node $i$.
Since node $i$ is connected with each node $j$ with probability $p_{ij}$ given by Eq. $(\ref{conn_p})$, the  probability $\mathcal{P}({\bf a}_i)$  is given by
\bea
\mathcal{P}({\bf a}_i)=\prod_{j=1}^N\left[p_{ij}a_{ij}+(1-p_{ij})(1-a_{ij})\right].
\eea
Using this result we can express the probability  $P(k_i|\theta_i,t_i)$ that node $i$ has degree $k_i$ as
\bea
P(k_i|\theta_i,t_i)&=&\sum_{{\bf a}_{i}}\mathcal{P}_i({\bf a}_{i})\delta\left(k_i,\sum_{j=1}^Na_{ij}\right)=\sum_{{\bf a}_{i}}\mathcal{P}({\bf a}_i)\int \frac{d\omega}{\sqrt{2\pi}}e^{-i\omega(k_i-\sum_{j=1}^Na_{ij})}
\eea
where we have used the integral representation of the Kronecker delta $\delta(x,y)$.
By performing the sum over all the  elements of ${\bf a}_i$ we get
\bea
P(k_i|\theta_i,t_i)&=&\int \frac{d\omega}{\sqrt{2\pi}}e^{-i\omega k_i}\prod_{j=1}^N\left[1-p_{ij}(1-e^{-i\omega})\right]=\int \frac{d\omega}{\sqrt{2\pi}} e^{F(\omega)}
\label{PF}
\eea
where
\bea
F(\omega)&=-i\omega k_i+\sum_{j=1}^N \ln\left[1-p_{ij}(1-e^{-i\omega})\right].\eea
For  $p_{ij}\ll1$ we can approximate $F(\omega)$ with
\bea
F(\omega)&=&i\omega k_i-\sum_{j=1}^Np_{ij}(1-e^{-i\omega})=i\omega k_i-\overline{k}_i(1-e^{-i\omega})\label{F}
\eea
where $\overline{k}_i$ is the expected degree of node $i$ given by
\bea
\overline{k}_i=\sum_{j=1}^N p_{ij}.
\label{ak}
\eea
Note here that since the connection probability $p_{ij}$ depends both on the hidden variables of the nodes $i$ and $j$ and on their arrival time in the network, it follows that also the expected degree $\overline{k}_i$ of node $i$  will be both a function of the {node's hidden variable and its} time of arrival in the network.
Using Eqs. (\ref{PF}) and (\ref{F}) we can derive  the explicit expression for $P(k_i|\theta_i,t_i)$. In fact we have
\bea
P(k_i|\theta_i,t_i)&\simeq &\int \frac{d\omega}{\sqrt{2\pi}} e^{i\omega k_i-\overline{k}_i(1-e^{-i\omega})}=\sum_{h=0}^{\infty}\frac{1}{h!}\overline{k}_i^h e^{-\overline{k}_i}\int \frac{d\omega}{\sqrt{2\pi}} e^{i\omega (k_i-h)},
\eea
and by identifying the last integral with the Kronecker delta $\delta({h,k_i})$ we get the Poisson distribution
\bea
P(k_i|\theta_i,t_i)=\frac{\overline{k}_i^{k_i}}{k_i!}e^{-\overline{k}_i}.
\eea
Therefore the probability that node $i${, which} arrived in the network at time $t_i$ with hidden variable $\theta_i${, } has degree $k_i$ is given by the Poisson distribution with average $\overline{k}_i$ given by Eq. (\ref{ak}).
It follows that the degree distribution $P(k)$ of the network at time $t$ is given by
\bea
P(k)=\int d\theta \rho(\theta)\frac{1}{t}\sum_{t'=1}^tP(k|\theta,t').
\eea
Note that for sufficiently sparse networks where each two connected nodes are typically connected by a link of weight one,  the degree of a node can be identified with its strength
\bea
k_i\simeq s_i.
\eea
{It} follows that  in this case the degree distribution can be approximated by the strength distribution and we have that if the hidden variables are power-law distributed with power-law $\gamma$ (as described in Eq. $(\ref{rhog})$) then also the degree distribution has a power-law tail with the same exponent $\gamma$, i.e.
\bea
P(k)=\tilde{C}k^{-\gamma}
\eea
for $k\gg 1$.

\subsection{Random permutation of the node sequence}

Here we  investigate  whether the described  network process can be related to the generation of uncorrelated networks.
In this way we aim at reconciling the non-equilibrium growing nature of the network model, displaying projectivity, with the properties of exchangeable but not projective uncorrelated network models.

We observe that this expression depends both on the hidden variable and on the time of arrival of the nodes $i$ and $j$ in the network.
However if we consider several realizations of the model in which the {times} of arrival of node $i$ and node $j$ {are} random, but the hidden variables are preserved, we observe that the probability that node $i$ and node $j$ are connected satisfies
\bea
P({a_{ij}=1|\theta_i,\theta_j,t=N})&=&\frac{1}{N^2}\int_1^N dt_i\int_1^N dt_j \int_1^N d\tau \delta(\tau, \min(t_i,t_j))\left[1-\exp\left(-\frac{\theta_i\theta_j}{\avg{\theta}\tau}\right)\right]\nonumber\\
&=&\frac{2}{N^2}\int_1^N d\tau \tau \left[1-\exp\left(-\frac{\theta_i\theta_j}{\avg{\theta}\tau}\right)\right]=2\frac{\theta_i \theta_j}{\avg{\theta}N}+o\left(\frac{\theta_i \theta_j}{\avg{\theta}N}\right).
\eea
Therefore if the network is sufficiently sparse, i.e.
\bea
\frac{\theta_i \theta_j}{\Avg{\theta} N}\ll 1,
\eea
we have that the expected degree $k_i(\theta_i)$ of a random node $i$ of hidden variable $\theta_i$ is given by
\bea
\tilde{k}_i(\theta_i)=2\theta_i,
\eea
and the probability that a node with hidden variable $\theta_i$ is connected with a node with hidden variable $\theta_j$ independently {of} their time of arrival in the network, is given by the uncorrelated network marginal corresponding to the number of nodes in the sample,
i.e.
\bea
\tilde{p}_{ij}=P({a_{ij}=1|\theta_i,\theta_j,t=N})= \frac{\tilde{k}_i(\theta_i) \tilde{k}_j(\theta_j)}{\avg{\tilde{k}(\theta)}N}.
\label{pij}
\eea
{Note} that in this case if the sample increases in size and includes $N^{\prime}>N$ nodes, the probability that node $i$ and node $j$ are connected will
satisfy
\bea
\tilde{p}_{ij}^{\prime}=P({a_{ij}=1|\theta_i,\theta_j,t=N^{\prime}})=\frac{\tilde{k}_i(\theta_i) \tilde{k}_j(\theta_j)}{\avg{\tilde{k}(\theta)}N^{\prime}}.
\eea
In this case  the network process induces a probability $\tilde{p}_{ij}$ that depends on the network size $N$ and at the same time enforces the sparseness of the network. In fact the expected degrees  $\{\tilde{k}_i\}$ of the nodes are only determined the the hidden variable and are independent on the network size. 

\subsection{Entropy of  the network model}

In order to compare our model  with hidden variable distribution $\rho(\theta)$ to an uncorrelated network ensemble in which the expected degrees are $\tilde{k}_i=2 \theta_i$, in this {section} we use information theory tools.
{  Specifically we } will compare the entropy of the two ensembles. The entropy of a network model or of a network ensemble \cite{Bianconi2007,Bianconi2009,Anand2009,Anand2010,PNAS} is a fundamental tool to evaluate the information content in the network model. It  indicates the logarithm of the typical number of networks generated by the ensemble and as such {evaluates} the complexity of the model and can be used in inference problems \cite{PNAS}.
Since for our network model the  connection probability $p_{ij}$ of any two pair of nodes is $i$ and $j$ is given by Eq. $(\ref{conn_p})$, the entropy of the model is given by
\bea
S=-\sum_{i<j}\left[p_{ij}\ln p_{ij}+(1-p_{ij})\ln(1-p_{ij})\right].
\eea
where in the sparse regime we can approximate $p_{ij}$ with $t_j>t_i$ as
\bea
p_{ij}\simeq\frac{\theta_i\theta_j}{\avg{\theta}t_j}.
\eea
Similarly for the uncorrelated network ensemble with connection probability $\tilde{p}_{ij}$   the entropy is given by
\bea
\tilde{S}=-\sum_{i<j}\left[\tilde{p}_{ij}\ln \tilde{p}_{ij}+(1-\tilde{p}_{ij})\ln(1-\tilde{p}_{ij})\right].
\eea
In order to compare these two entropies we use the explicit expression for the connection probability $\tilde{p}_{ij}$ when we put $\tilde{k}_i(\theta_i)=2\theta_i$ which reads
\bea
\tilde{p}_{ij}=2\frac{\theta_i\theta_j}{\avg{\theta}N}.
\eea
By performing a straightforward calculation we find that $S$ is given, up to the linear terms in $N$, by
\bea
{  S=\avg{\theta} \ln (N!)-2N \avg{\theta\ln\theta}+N\avg{\theta}\ln \avg{\theta}+\avg{\theta}N}
\eea
and that  the entropy $S$ of our model is { smaller} than  the entropy of the uncorrelated network ensemble. In fact, $S$ differs from $\tilde{S}$  only by
\bea
{ \Delta S=S-\tilde{S}\simeq \avg{\theta}\ln \left(\frac{N!2^N}{N^N}\right)\simeq\avg{\theta}N\ln\left(\frac{2}{e}\right).}
\eea
{   The entropy difference  $\Delta S$ quantifies  the  information loss when the proposed network process is approximated  with its corresponding  uncorrelated network model.
We observe here that the uncorrelated network model is obtained when  the causal construction of the original network model is disregarded  and  the only retained information is  the probability $\tilde{p}_{ij}$ that two nodes of hidden variables $\theta_i$ and $\theta_j$ are connected regardless of their time of arrival in the network. Therefore $\Delta S$ captures the loss of information when the causal nature of the original model is disregarded.
Interestingly in the large network limit $N\gg 1$, $|\Delta S|$ is low when compared to $S$ revealing the proximity between the two models. Additionally $\Delta S$  is only dependent on $\Avg{\theta}$ indicating that the information loss from one model to the other  is independent of   the particular distribution of the hidden variables $\rho(\theta)$ as long as  $\Avg{\theta}$ is kept constant.}

\section{Statistical testing of the model}
\label{Stat}

\begin{figure}[!htb]
  \noindent \begin{centering} \includegraphics[scale=0.65]{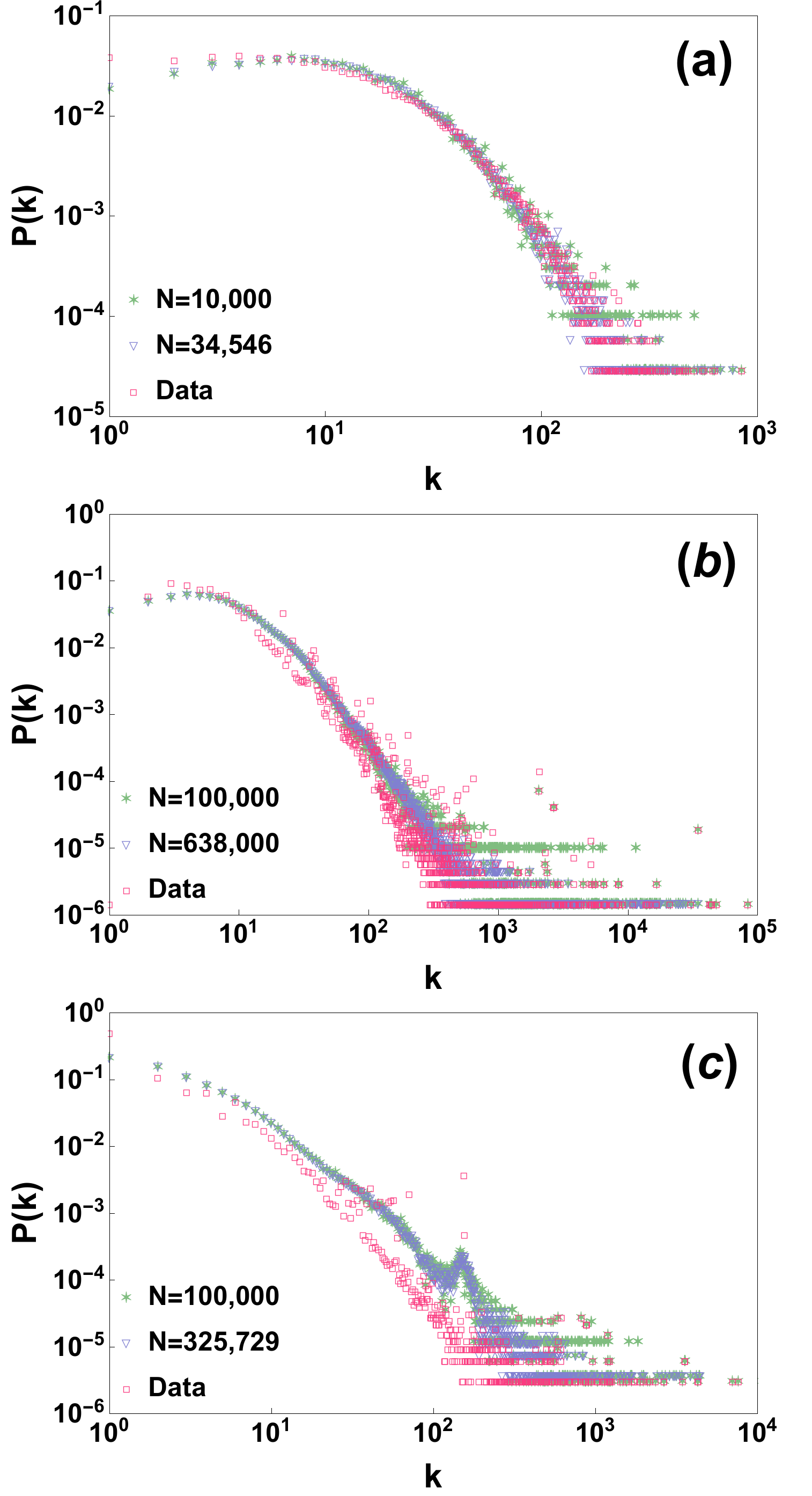}
 \par\end{centering}
\caption{The degree distributions $P(k)$ of the three analysed datasets is compared with the results of the model generated by using  all the nodes of the network or with just a subsample of nodes of the network of size $N$. Panels (a,b,c) display the results for  the arxiv hep-ph  citation network \cite{citations1,citations2} ($N=34,546$) the Berkeley-Stanford web network \cite{BS_WWW} ($N=685,546$) and the Notre Dame web network \cite{ND} ($N=325,000$) respectively.}
\label{fig1}
\end{figure}

In order to study the utility of the proposed model {as a null model for} sampled data we consider three power-law networks:  the arxiv hep-ph (high energy physics phenomenology) citation network \cite{citations1,citations2}, the Berkeley-Stanford web network \cite{BS_WWW} and the Notre Dame web network \cite{ND} of network sizes $N=34,546$, $N=685,230${,} $N=325,000$ respectively. All data are freely available on the Stanford Network Analysis Project webpage.
To each node of the network we assign a different label $i\in{1,2,\ldots,N}$ according to  a random permutation of the indices from $1$ up to $N$.
We {then assign} to each node $i$ of the network {a hidden} variable
\bea
\theta_i=\frac{1}{2}k_i{,}
\eea
where $k_i$ is the observed degree of node $i$ in the dataset.
Given our random node labelling and the hidden variables $\{\theta_i\}_{i=1,2,\ldots,N}$ we have generated a random network according to the proposed network process.
Interestingly the proposed  model preserves to {a} large extent the degree distribution (see comparison of the real degree distribution with the one generated by the model in Figure $\ref{fig1}$).
Additionally these results are {quite} stable if we  consider a model generated only by adding a subsample of  randomly chosen nodes, showing that the model preserves the degree distribution under random sub-sampling of the nodes (see Figure $\ref{fig1}$).

\begin{figure}[!htb]
  \noindent \begin{centering} \includegraphics[scale=0.65]{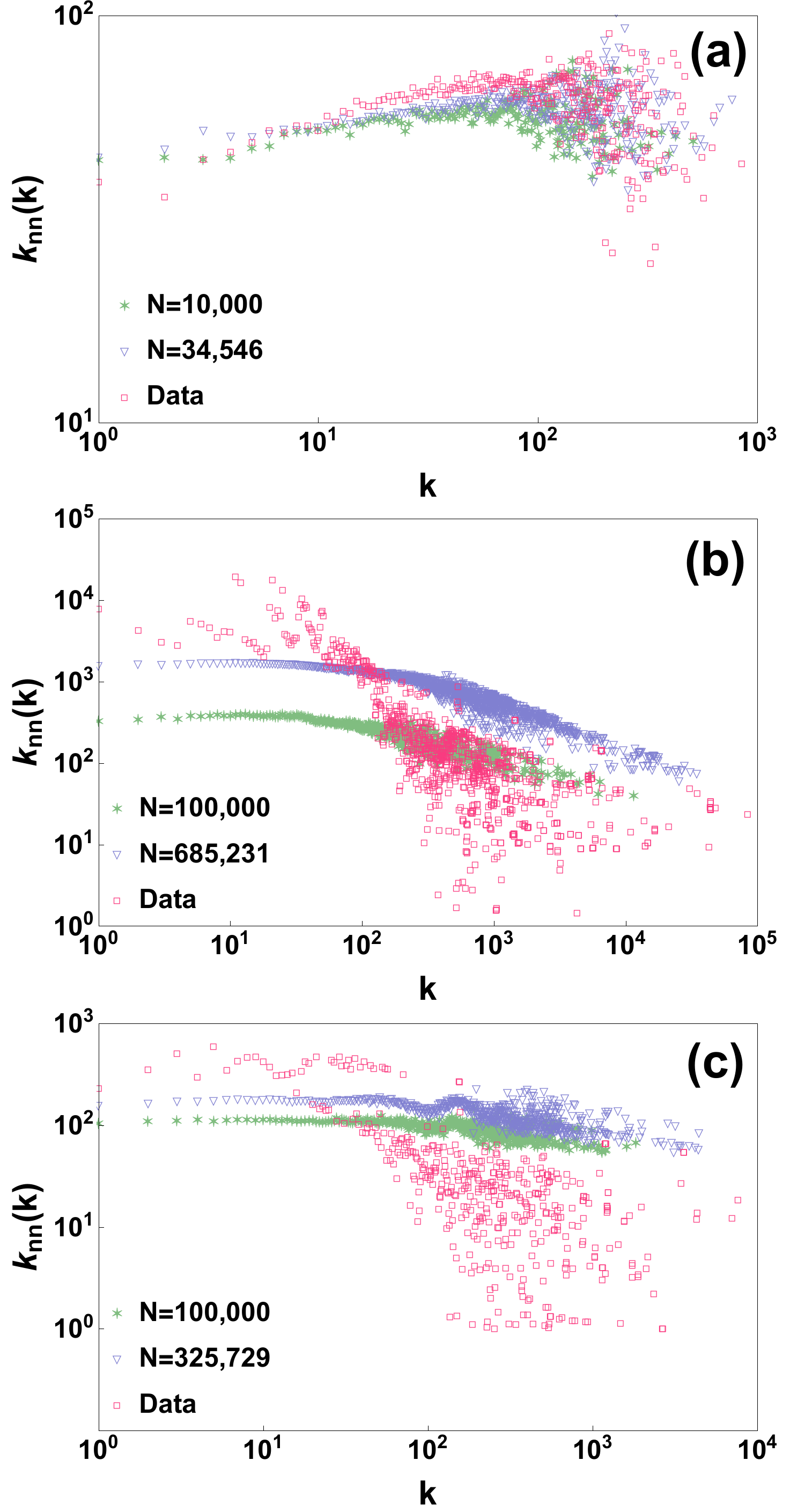}
\par\end{centering}
\caption{The average degree $k_{nn}(k)$ of the neighbour of a node of degree $k$ of the three analysed datasets is compared with the results of the model generated by using  all the nodes of the network or with just a subsample of nodes of the network of size $N$. Panels (a,b,c) display the results for  the arxiv hep-ph  citation network \cite{citations1,citations2} ($N=34,546$) the Berkeley-Stanford web network \cite{BS_WWW} ($N=685,546$) and the Notre Dame web network \cite{ND} ($N=325,000$) respectively.} \vspace{4mm}
\label{fig2}
\end{figure}

The generated model however  is to be considered mostly as  uncorrelated. In fact if we compare the degree correlations of the real datasets with the degree correlations of the network generated by the model we observe that the model deviates from the real data and {displays} very weak/marginal degree correlations (see Figure $\ref{fig2}$). {  In fact  from the results obtained for the three studied network datasets it seems that the model is able to better reproduce weakly assortative behaviour than strongly disassortative bahaviour.} In future,  modifications of the proposed model could be envisaged to capture also the degree correlations of real datasets.

 \section{Conclusions}

In conclusion{,} we have given a wide overview of {the desirability of the projectivity and exchangeability properties in good statistical models} and we have emphasized the difficulty in combining these properties with the sparseness of the network.
While this {problem} is a widely discussed subject in statistics of networks and graph theory, here we have proposed a model that provides a trade-off solution.
Our model  describes a network process in which nodes and links are subsequently added according to a probability dependent {on} some hidden variables associated to the nodes.
As long as the hidden {variables} are power-law distributed this model  generates a scale-free network with the same exponent. This model  is projective but {not }exchangeable. However{,} the expected probability that two nodes are connected when one consider{s} a random permutation of the  sequence in which nodes are  added to the network  reduces to the expression valid for the marginal {of} an uncorrelated exchangeable network with the same expected degrees {(given by the double of the hidden variables)} provided the network is sufficiently sparse.
{Finally, we} tested this model as a statistical null model for scale-free sparse real network{s}, showing that it can reproduce the degree distribution {(but not degree correlations)} also if a partial subset of the data is considered.

\section*{Acknowledgements} D.K. acknowledges funding from grant numbers ARO W911NF-16-1-0391 and NSF IIS-1741355.  J.P.G. acknowledges funding from Science Foundation Ireland (grant number 16/IA/4470). Y. M. acknowledges partial support from the Government of Arag\'on, Spain through a grant to the group FENOL, and by MINECO and FEDER funds (grant FIS2014-55867-P).

\bibliography{bib_final}

\begin{thebibliography}{48}
\bibitem[{spa()}]{sparse_note}
\bibinfo{note}{We note here that while in the statistics literature the term
  {\it sparse network} refers to a network where the average degree is
  sublinear in the number of nodes $N$, i.e. $\avg{k}=o(N)$ here we adopt the
  physics terminology and use the term {\it sparse network} to indicate
  networks with average degree independent of $N$, i.e. $\avg{k}=O(1)$. These
  networks are also indicated in the statistical literature as {\it
  ultra-sparse}.}
\end{thebibliography}

\end{document}